\documentclass[aps,prd,showpacs,nofootinbib,superscriptaddress,preprint,tightenlines]{revtex4}

\usepackage{graphicx}
\usepackage[centertags]{amsmath}
\usepackage{amsfonts}
\usepackage{amssymb}
\usepackage{amsthm}
\usepackage{newlfont}
\usepackage{subfigure}
\usepackage{multirow}
\usepackage{accents}
\usepackage{color}
\usepackage{float}

\def\bea{\begin{eqnarray}}
\def\eea{\end{eqnarray}}
\def\be{\begin{equation}}
\def\ee{\end{equation}}

\usepackage{graphicx}

\def\be{\nopagebreak[3]\begin{equation}}
\def\ee{\end{equation}}
\def\ba{\nopagebreak[3]\begin{eqnarray}}
\def\ea{\end{eqnarray}}

\newcommand{\teta}{\rlap{\lower2ex\hbox{$\,\tilde{}$}}\eta{}}

\def \Gc{\Gamma_{\text{cov}}}
\def \cGc{\bar{\Gamma}_{\text{can}}}

\newcommand{\mr}{\mathrm}

\usepackage{enumerate}

\usepackage{colordvi}

\usepackage[centertags]{amsmath}
\usepackage{amsfonts}
\usepackage{amssymb}
\usepackage{amsthm}
\usepackage{dsfont}
\usepackage{newlfont}
\usepackage{hyperref}
\def\be{\begin{equation}}
\def\ee{\end{equation}}

\newcommand{\md}{{\mathrm d}}

\newcommand{\ed}{\md \!\!\!\! \md\, }

\begin{document}

\title{On covariant and canonical Hamiltonian formalisms for gauge theories}
\author{Alejandro Corichi}\email{corichi@matmor.unam.mx}
\affiliation{Centro de Ciencias Matem\'aticas, Universidad Nacional Aut\'onoma de
M\'exico, UNAM-Campus Morelia, A. Postal 61-3, Morelia, Michoac\'an 58090,
Mexico}

\author{Juan D. Reyes}
\email{jdreyes@uach.mx}
\affiliation{Facultad de Ingenier\'\i a,
Universidad Aut\'onoma de Chihuahua, 
Nuevo Campus Universitario, Chihuahua 31125, Mexico}

\author{Tatjana Vuka\v{s}inac}
\email{tatjana.vukasinac@umich.mx}
\affiliation{Facultad de Ingenier\'\i a Civil, Universidad Michoacana de San Nicol\'as de Hidalgo, 
Morelia, Michoac\'an 58000, Mexico}

\begin{abstract}
The Hamiltonian description of classical gauge theories is a very well studied subject. The two best known approaches, namely the covariant and canonical Hamiltonian formalisms have received a lot of attention in the literature. However, a full understanding of the relation between them is not available, specially when the gauge theories are defined over regions with boundaries. Here we consider this issue, by first making precise what we mean by equivalence between the two formalisms. Then we explore several first order gauge theories, and assess whether their corresponding descriptions satisfy the notion of equivalence. We shall show that, 
even when in several cases the two formalisms are indeed equivalent, there are counterexamples that signal that this is not always the case. Thus,  non-equivalence is  a  generic feature for gauge field theories. These results call for a deeper understanding of the subject.
\end{abstract}

\pacs{03.50.Kk, 11.15.Yc, 11.10.Ef}
\maketitle


\section{Introduction}
\label{sec:1}

Gauge theories defined on spacetime regions with boundary can have degrees of freedom and observables localized on the boundary. One can study them following one of (at least) two approaches to a Hamiltonian formalism: Covariant and canonical methods. It is a reasonable expectation that, at the classical level, those two descriptions should be equivalent. There are several papers that have dealt with this issue (see for instance \cite{Wald,Espanolitos}), where the claim is made that both methods are equivalent. This question is of great importance and should be considered also when there are boundaries present, since the derived Hamiltonian descriptions  may result in inequivalent physical predictions. The purpose of this manuscript is to revisit this issue, for simple cases of (first order) gauge theories defined over regions with boundaries, where both covariant and canonical methods are well understood. As we shall see, there are nontrivial examples for which such assumed equivalence is at odds with already known results. Thus, we shall point out some of such instances, in the hope that more thorough investigations shall fully clarify the issues at hand.

The first question one might ask is how the two descriptions and in particular, the symplectic structure (fundamental for the definition of dynamics), can be equivalent when the corresponding phase spaces are {\it different}  objects. To be precise, in the canonical approach, the phase space $\Gamma_{\mathrm{can}}$ is given by initial data on a hypersurface $\Sigma$, and the (so called kinematical) symplectic structure is the `canonical' one on the 
cotangent bundle over the configuration space (more below). On the other hand, the covariant Hamiltonian formalism is based on the covariant phase space $\Gamma_{\mathrm{cov}}$ defined by the space of solutions to the equations of motion. Are there instances where a diffeomorphism between these two spaces can be defined? The answer is in the affirmative, for some systems. Consider, for instance, a scalar field satisfying a linear equation of motion on an arbitrary, globally hyperbolic spacetime. If the corresponding fields are appropriately well behaved, one might have existence and uniqueness of solutions to the equations of motion, given initial data. In that case, one can define a mapping $\mathcal{I}: \Gamma_{\mathrm{can}} \rightarrow \Gamma_{\mathrm{cov}}$. It is invertible, so we can define all kinds of 1-1 mappings between objects, such as the symplectic structure, defined on both spaces \cite{Wald-book}. 

When we consider Hamiltonian gauge theories, namely singular systems with so called first class constraints (FCC), then one immediately runs into problems. The first issue is that a bijection is lost. Given a solution to the covariant equations of motion, one can induce initial data on (a preferred) hypersurface $\Sigma_0$, that satisfies the constraints of the canonical theory. If we now evolve that initial data, there will not be a unique solution, precisely due to the freedom of adding arbitrary linear combinations of the constraints to the Hamiltonian. Thus we cannot define a 1-1 map between $\Gamma_{\mathrm{cov}}$ and the constraint surface  $\bar{\Gamma}_{\mathrm{can}}$.
All this should be such that, when one quotients out the gauge orbits, and arrives at the corresponding reduced phase space for each approach, then there shall be complete equivalence between them. That is, at the level of reduced phase spaces, one expects them to be fully equivalent.
The problem is that, in practice, it is very difficult to work in the reduced phase space (one not even has control over it, in generic cases). So the best one can do is to obtain, at the level of pre-symplectic spaces,  similar expressions 
for the pre-symplectic structures. This is precisely the issue that we shall examine here. It is, of course, important to  be clear about what one is comparing, so we shall make precise what we mean by equivalence further below.

We present some preliminaries regarding both formalisms in Sec. \ref{sec:2}, and we state precisely the notion of equivalence that we shall consider. Next, we explore in detail several simple gauge theories, in regions with boundaries. In Sec. \ref{sec:2.5}, we start with one of the better known gauge theories in 4 dimensions,  corresponding to the Maxwell field. Here it is well known that, given the linearity of the theory, both canonical and covariant descriptions are equivalent (see for instance \cite{Corichi1998} for such an analysis). We shall provide a brief review of such theory.
The next system we shall consider is a topological theory, namely $U(1)$ Pontryagin in the bulk. In Sec. \ref{sec:3} we shall analyze both the covariant and the canonical approaches and show how different the descriptions are. At the end of the day, of course, we arrive to a description which is `trivial', with no local physical degrees of freedom. 
The next system we will consider in Sec. \ref{sec:4} is a $U(1)$ Chern-Simons theory on the boundary (related to the Pontryagin theory). It is known that, at the level of the action, both theories are equivalent. It has also been shown that, when considering  the canonical phase space, there is also equivalence in a precise sense \cite{CV-P}. In Sec.\ref{sec:5}, we shall analyse Maxwell-Pontryagin in both approaches. Here we see that they are not equivalent in the sense we shall describe below. To be specific, the two  structures, at the pre-symplectic level, differ by a boundary term. Finally, in Sec.\ref{sec:6} we consider Maxwell in the bulk with Chern-Simons term on the boundary. We see in this case that the pre-symplectic structures are also equivalent. We hope that these results shall add to  our collective understanding of the relation between  canonical and covariant Hamiltonian methods.

Throughout the manuscript, we are considering generic globally-hyperbolic spacetimes without any specific choice of metric, nor coordinate system. Furthermore, we employ the language of forms, since that simplifies the calculations and allows for shorter expressions. When needed, we have re-written previously known results to match this notation. 

This contribution is to appear in a volume in honor to Prof. Ashtekar, who has been a pioneer in the application of both canonical \cite{Ashtekar, Abhay1} and covariant approaches \cite{abr, afk} to general relativity. 


\section{Preliminaries on covariant and canonical Hamiltonian analysis}
\label{sec:2}

In this section we will give a very short reminder of the basic ideas of both approaches, based mostly on \cite{witten,abr,Wald,CVZ2} for the covariant case, and \cite{HT,CV-M+P,GNH,Barbero} for the canonical one, among many others.
For simplicity and concreteness here we shall consider first order gauge theories, whose configuration space is formed from $U(1)$ connection $1-$forms $\mathbf{A}(x)$ given on a spacetime region $\mathcal{M}$ with boundary, $\partial\mathcal{M}=\Sigma_1\cup\Sigma_2\cup\mathcal{B}$, where $\Sigma_{1,2}$ are two (arbitrary) Cauchy surfaces and $\mathcal{B}$ is a time-like hypersurface. The connections should satisfy some appropriate boundary conditions that can be either given a priori, from some (physical) considerations, or can be obtained in the process of  constructing a consistent theory, or the combination of both. 
Here we want to compare the two approaches, the covariant and the canonical ones. Let us first recall both formalisms. 

The covariant phase space $(\Gamma_{\text{cov}},\pmb{\omega})$,  consists of $\Gamma_{\text{cov}}$, the space of solutions to the equations of motion, that satisfy some appropriate boundary conditions, together with a pre-symplectic structure $\pmb{\omega}$, that is a degenerate, closed $2-$form on $\Gamma_{\text{cov}}$. 
The degenerate directions $Z_i$, are such that $\pmb{\omega} (Y, Z_i)=0$, for every $Y\in T\Gamma_{\text{cov}}$. 
The degeneracy signals  the existence of gauge orbits in $\Gamma_{\text{cov}}$, that relate physically equivalent states. The space of such orbits ${\hat\Gamma}_{\text{cov}}$, corresponding to  different physical states, is the so called reduced phase space, where the projection $\hat{\pmb{\omega}}$ under the quotient map is non-degenerate.   

On the other hand, the canonical phase space $(\Gamma_{\text{can}},\Omega)$ is the space of all allowed initial data $\Gamma_{\text{can}}$ equipped with a
non-degenerate, closed $2-$form $\Omega$, the kinematical symplectic structure.
In gauge theories there are first class constraints and when we restrict to the constraint surface, the pullback of the symplectic structure $\bar{\Omega}$ becomes degenerate. Then, in principle, one could solve the constraints and impose some gauge fixing conditions (if possible), in order to obtain a reduced phase space with a non-degenerate symplectic structure $\hat{\Omega}$.


In both cases the starting point is a covariant action. In the covariant analysis, we find the geometric structures and arrive to a symplectic description in a natural way without the need to foliate the underlying spacetime region, nor project the fields to some hypersurfaces.
In the canonical case, one has to perform a decomposition of the spacetime region, of the fields and of the covariant action to arrive at a canonical one. 
Since $\mathcal{M}$ has a boundary, both symplectic structures can have a boundary contribution, but, as it turns out, they do not necessarily coincide. In the following we shall recall both constructions.





\subsection{Covariant Hamiltonian analysis}
\label{subsec:2.1}



Let us start from a generic,  first order covariant action, without boundary terms,
\begin{equation}
S[\mathbf{A}] = \int_{\mathcal{M}}  \mathbf{L}\, .\label{CovAction}
\end{equation}
Its variation can be written as 
\begin{equation}\label{VarActFormsWithoutBoundary}
\delta S [\mathbf{A}]   = \int_{\mathcal{M}} \mathbf{E} \wedge\delta \mathbf{A}  + \int_{\mathcal{M}} \md \theta ( \mathbf{A},\delta \mathbf{A}).
\end{equation}
The second term of the RHS is obtained after integration by parts. If it vanishes, we obtain the Euler-Lagrange equations of motion,  $\mathbf{E}=0$, in the bulk. Though its vanishing is a condition needed for the well defined action principle, this term, called the symplectic potential, is also a starting point for the construction of a symplectic structure of the theory. Using Stokes' theorem it can be written as,
\begin{equation}\label{SympPotentialWhitoutBoundary}
\Theta (\delta \mathbf{A}) := 
\int_{\partial \mathcal{M}} \theta (\mathbf{A},\delta \mathbf{A} )\, .
\end{equation}
The exterior derivative of the symplectic potential, 
acting on tangent vectors $\delta_{1}$ and $\delta_{2}$ at a point $s \in \Gamma_{\text{cov}}$ is given by
\begin{equation}\label{edTheta}
\ed \Theta (\delta_{1}, \delta_{2}) := \delta_{1} \Theta (\delta_{2}) -\delta_{2} \Theta (\delta_{1}) = 2 \int_{\partial \mathcal{M}} \delta_{[1} 
\theta (\delta_{2]})\, ,
\end{equation}
where $\ed$ denotes the exterior derivative in the phase space, and where one identifies variations of the fields with tangent vectors to the space 
$\Gamma_{\text{cov}}$.
From this expression we can define a space-time $3-$form, the symplectic current $J(\delta_{1}, \delta_{2})$, as
\begin{equation}\label{DefJ}
J(\delta_{1}, \delta_{2}) := \delta_{1} \theta  (\delta_{2}) -\delta_{2} \theta (\delta_{1})\, .
\end{equation}
On the space of solutions, $\ed S (\delta) = \Theta (\delta)$, therefore we obtain
\begin{equation}
0 = \ed ^{2} S (\delta_{1}, \delta_{2}) =  \ed \Theta (\delta_{1}, \delta_{2}) = \left(-  \int_{\Sigma_{1}} + \int_{\Sigma_{2}}    + 
\int_{\mathcal{B}} \right)  J \, .
\end{equation}
In some examples the 
boundary conditions ensures that the integral $\int_{\mathcal{B}}  J$  vanishes, in that case it follows that
$\int_{\Sigma}  J$ is independent of the Cauchy surface. This allows us to define a \emph{conserved} pre-symplectic form over an arbitrary space-like Cauchy surface $\Sigma$,
\begin{equation}
\pmb{\omega} (\delta_{1}, \delta_{2}) = \int_{\Sigma}  J  (\delta_{1}, \delta_{2})\, .
\end{equation}
In this case there is no boundary contribution in $\pmb{\omega}$.

The boundary term appears when $J=\md j$ on $\mathcal{B}$. Then, $\int_{\mathcal{B}}  J= (-\int_{\partial\Sigma_1}+\int_{\partial\Sigma_2}) j$, and the conserved pre-symplectic
structure takes the form
\begin{equation}
\pmb{\omega} (\delta_{1}, \delta_{2}) = \int_{\Sigma}  J  (\delta_{1}, \delta_{2})+\int_{\partial\Sigma}  j  (\delta_{1}, \delta_{2}) \, .
\end{equation}
In the following we shall revise various examples of the theories where the covariant action has an additional topological term. Let  us recall why this term does not affect the covariant symplectic structure. In this case we have 
\begin{equation}\label{ActionFormsWithBoundary}
S[\mathbf{A}] = \int_{\mathcal{M}}  \mathbf{L} + \int_{\mathcal{M}} \md \Phi \, .
\end{equation}
On the equation of motion we obtain
\begin{equation}\label{GeneralVariatonForms}
\delta S[\mathbf{A}] =  \int_{\mathcal{M}} \md \left[\theta (\mathbf{A},\delta \mathbf{A}) +  \delta  (\Phi (\mathbf{A})) \right]\, .
\end{equation}
and the symplectic potential now has an additional term 
\begin{equation}
\tilde \Theta (\mathbf{A},\delta \mathbf{A}) := 
\int_{\partial \mathcal{M}}  \left[\theta (\mathbf{A}, \delta \mathbf{A}) +  \delta  (\Phi (\mathbf{A})) \right]\, .\label{sympl_poten_boundary_term}
\end{equation}
From (\ref{sympl_poten_boundary_term}) it follows that the corresponding symplectic current is of the form
\begin{equation}\label{defJ}
\tilde {J}(\delta_{1}, \delta_{2}) = 2 \bigl(\delta_{[1} \theta (\delta_{2]}) + \delta_{[1} \delta_{2]} \Phi 
 \bigr)\, .
\end{equation}
Now, we see that the term $ \delta_{[1} \delta_{2]} \Phi $ vanishes by antisymmetry.
Therefore, when we add a topological term to the original action it will not change the symplectic current, and, as a consequence, neither the symplectic structure of the original theory \cite{CVZ2}.

In the covariant phase space framework, the energy of the system, $H$ is determined (up to an additive constant) from
\be
\delta H :=\ed H (\delta ) = \pmb{\omega}(\delta , \delta_t)\, ,
\ee 
where  $\delta_t\mathbf{A}={\mathcal{L}}_t\mathbf{A}$, and $\delta \in T\Gamma_{\text{cov}}$. 

It is interesting that in diffeomorphism invariant theories, such as the theory of gravity in first order fomalism, $\pmb{\omega}(\delta_1 , \delta_2)={\pmb{\omega}}_{\text{bulk}}(\delta_1,\delta_2)+{\pmb{\omega}}_{\text{bound}}(\delta_1,\delta_2)$. It turns out that in the asymptotic region (for asymptotically flat configurations) and on the (weakly) isolated horizon, as an internal boundary, ${\pmb{\omega}}_{\text{bulk}}(\delta , \delta_t)$ reduces to an integral over a 2D surface 
$\partial\Sigma$, while ${\pmb{\omega}}_{\text{bound}}(\delta , \delta_t)$ vanishes. 
As a consequence $H$ is determined as an integral over a 2D surface 
$\partial\Sigma$, see, for example, \cite{afk}.


\subsection{Canonical Hamiltonian analysis}
\label{subsec:2.2}
In this part, we shall  briefly recall the procedure to arrive at the canonical phase space $\Gamma_{\mathrm{can}}$.
The starting point is again a covariant action (\ref{CovAction}) or (\ref{ActionFormsWithBoundary}), with spacetime region 
$\mathcal{M}$ of the form $\mathcal{I} \times  \Sigma$, with $ \mathcal{I}$ a closed interval.
The 3+1 decomposition of the action, through a spacetime foliation, and a choice of time evolution vector field, 
amounts to a field redefinition or change of variables in configuration space, which allows to write the action in canonical form:
\be
S_{\text{can}}
=\int_\mathcal{I} ( P[{\mathcal{L}}_t \mathbf{A}] - H_{\text{C}} )\, \md t\, ,
\ee
where $H_{\text{C}}$ is the (candidate for)  canonical Hamiltonian of the theory (that can also have a boundary term) and $P[{\cal{L}}_t\mathbf{A}]$ is the kinetic term, that can also have a contribution from the boundary \cite{CV-M+P},
\be
P[{\mathcal{L}}_t {\mathbf{A}}] = \int_\Sigma\,  \mathbf{P} \wedge {\mathcal{L}}_t \mathbf{A} + 
\int_{\partial\Sigma}\, \pmb{\pi}\wedge {\mathcal{L}}_t \pmb{\alpha}\, . 
\ee
Here $(\mathbf{A}(x),\mathbf{P}(x))$ are the  bulk canonical variables and
$(\pmb{\alpha}(y),\pmb{\pi}(y))$ are interpreted as boundary degrees of freedom.
In general, boundary fields are not the pullback of bulk phase space variables to $\partial\Sigma$. Thus, one has obtained coordinates for the cotangent bundle $T^*\cal{C}=:\Gamma_{\mathrm{can}}$, where $(\mathbf{A},\pmb\alpha)$ are coordinates for the canonical configuration space in the bulk, and boundary, respectively. Furthermore, $(\mathbf{P},\pmb{\pi})$ are  the corresponding momenta over $\cal{C}$.

The kinetic term determines the kinematical non-degenerate symplectic structure of the theory that can also have a boundary term\footnote{More specifically, assuming well defined tangent and cotangent bundle structures for the configuration space $\mathcal{F}$, $P[{\cal{L}}_t\mathbf{A}]$ allows one to identify the symplectic potential and symplectic structure determined by the Euler-Lagrange equations in the tangent space $T\mathcal{F}$,  and to map them through the Legendre transform to the canonical 1- and 2- forms on the cotangent bundle $T^*\mathcal{C}$.}
\be
\Omega (\delta_1,\delta_2)= 2\int_\Sigma \delta_{[1}\mathbf{P}\wedge\delta_{2]}\mathbf{A}   + 2\int_{\partial\Sigma} \delta_{[1}\pmb{\pi}\wedge\delta_{2]}\pmb{\alpha}\, .
\ee
As mentioned before, in gauge theories there are first class constraints $\mathbf{C}_i\approx 0$, $i=1,\dots,n$, and in general, there can be bulk and boundary ones. The theory can also have second class constraints (SCC), $\mathbf{D}_k\approx 0$, $k=1,\dots,m$. For our purposes, it is sufficient to restrict our considerations to the subspace of the phase space where the SCC constraints are imposed as strong equalities, $\Gamma_D\subset\Gamma_{\text{can}}$.

The evolution is tangent to the first class constraint surface,
$\bar{\Gamma}_{\text{can}}\subset\Gamma_{\text{can}}$ and the  pullback of $\Omega$ to $\bar{\Gamma}_{\text{can}}$ becomes degenerate. The Hamiltonian that governs the dynamics of the theory is  obtained by adding to the canonical Hamiltonian a linear combination of the smeared FCC constraints $H=H_{\text{C}}+{\mathbf{C}_i}[u_i] $, where $u_i$ are arbitrary multipliers.  
For any choice of the multipliers, there is a unique  corresponding Hamiltonian vector field $X_H$ defined as
\be
\ed H (Y)= \Omega (Y ,X_H)\, ,\label{HVF1}
\ee
where $Y\in T\Gamma_{\text{can}}$. The vector field $X_H$ should be tangent to $\bar{\Gamma}_{\text{can}}$. Let us denote by $Z_i$ the Hamiltonian vector fields that correspond to first class constraints, then
\be
\bar{\Omega} (Z_i ,X_H)=0\, , \ \ i=1,\dots ,n\, , \label{HVF2}
\ee
where $\bar{\Omega}$ is the pullback of $\Omega$ to $\bar{\Gamma}_{\text{can}}$.


Even when one has the same Hamilton equations in both formalisms, there is a subtle difference in how one uses them. Contrary to the covariant formalism, in the canonical approach one starts from $H$ and constructs the corresponding $X_H$. There are two possibilities depending on the form of the symplectic structure. In the first one $\Omega$ has vanishing contribution from the boundary so there cannot be any boundary terms in (\ref{HVF1}) and (\ref{HVF2}), and that imposes some boundary conditions on bulk canonical variables.  This corresponds to the standard Regge-Teitelboim scenario. 
In the second case $\Omega$ has non vanishing contribution from the boundary. Then, there is a boundary contribution to Hamiltonian vector fields, and generally there are also boundary conditions on the bulk configurations \cite{CV-M+P}.  

Let us now compare both formalisms and recall how they are related, and how one can pose the problem of having equivalent structures at the pre-symplectic level.

\subsection{Comparison and statement of the problem}

In the introduction we saw that one can define a mapping from the space of solutions to the constraint surface in the canonical phase space.  The 3+1 decomposition provides a `canonical' way to implement this map through a spacetime foliation,  assigning to each solution s in $\Gc$ its `instantaneous' value d at some `initial' hypersurface $\Sigma_0$. Let us call this map $\bar{\Pi}:\Gc\to\cGc$, the \emph{canonical projection}.
For gauge theories there are many solutions that induce the same initial data, so this map is non-injective. In analogy with the Lee-Wald construction \cite{Wald}, let us assume $\bar{\Pi}:\Gc\to\cGc$ gives $\Gc$ the structure of a fiber bundle over $\cGc$.

\begin{figure} 
     \begin{center}
      \includegraphics[width=9cm]{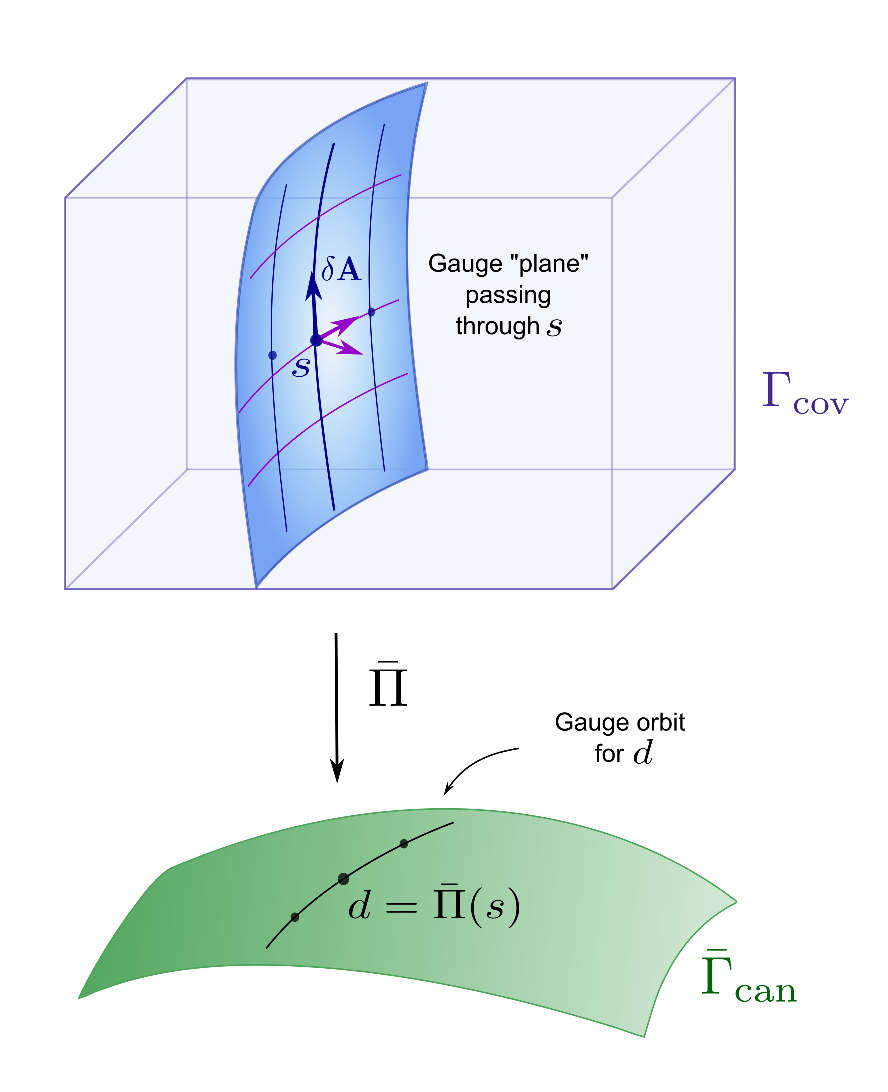}
     \caption{Gauge `plane' on $\Gamma_\text{cov}$: Directions $\delta\bold{A}$ along the fiber $\bar{\Pi}^{-1}(d)$ necessarily represent infinitesimal gauge transformations on $\mathcal{M}$ with support outside $\Sigma_0$. Transverse gauge directions on $\Gamma_\text{cov}$ (infinitesimal gauge transformations on $\mathcal{M}$ which are nontrivial on $\Sigma_0$) should connect fibers over the gauge orbit of $d$ in the constraint surface $\bar{\Gamma}_{\text{can}}$. Note that there are several transverse vectors at $s$ that project to the same vector in $d$.} \label{f1}
      \end{center}
\end{figure}

As we have seen, one expects that all those different solutions mapping to the same initial data  $d$, that is, all the points along the fiber  $\bar{\Pi}^{-1}(d)$, belong to the same gauge orbit in $\Gc$. There are, furthermore, the gauge orbits in $\cGc$ generated by the constraints. Those points along the orbit of $d$ in $\cGc$, associated to initial data for different solutions, will have their corresponding fibers over them. Those fibers should also belong to the same gauge orbit in $\Gc$ as the initial fiber we started with. Thus, intuitively, the gauge orbits in $\Gc$ shall be larger (of higher dimension per point) than the gauge orbits in $\cGc$ (See Fig. \ref{f1})

\begin{figure}  
     \begin{center}
      \includegraphics[width=9cm]{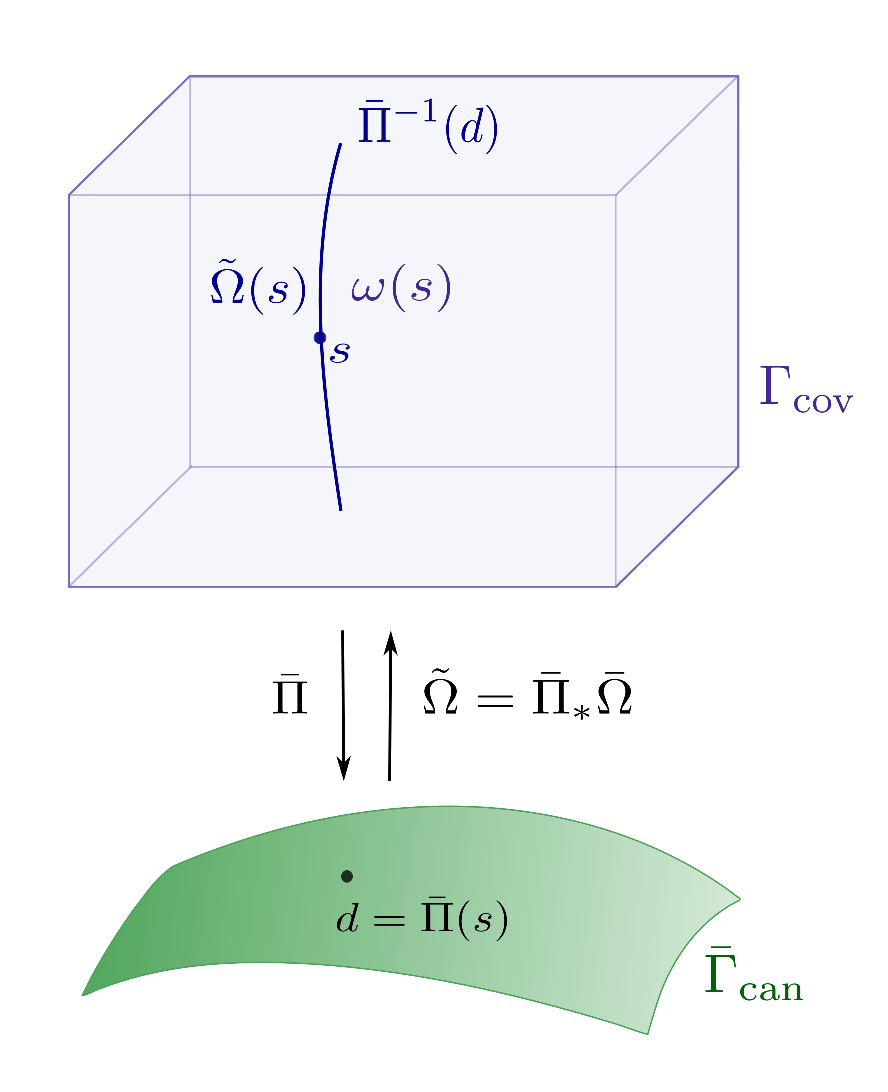}
      \caption{$\tilde{\Omega}$, the pullback under the canonical projection map of the pre-symplectic structure $\bar{\Omega}$  on the constraint surface,  defines a pre-symplectic structure on the covariant phase space. Directions $\delta\bold{A}$ along the fiber $\bar{\Pi}^{-1}(d)$ are degenerate directions of $\tilde{\Omega}(s)$. Since $\tilde{\Omega}$ is closed, by Cartan's formula then $\mathcal{L}_{\delta\bold{A}}\tilde{\Omega}=0$. So $\tilde{\Omega}$ also has a well defined projection.}  \label{f2}
           \end{center}
\end{figure}

The question that was left open in the introduction was a precise way of defining an equivalence. In the discussion we argued that this can be done at the level of pre-symplectic spaces, that is, relating $\Gc$ and $\cGc$. Let us see that in detail now. Using the canonical projection $\bar{\Pi}(s) = d$, with $s$ a solution and $d$ its corresponding initial data, 
the pullback $\tilde{\Omega}(s) :=\bar{\Pi}_*\bar{\Omega}(d)$ of the pre-symplectic structure $\bar{\Omega}$ at each point $d$ on $\bar{\Gamma}_\text{can}$, defines a pre-symplectic structure on $\Gamma_\text{cov}$ (a closed 2-form).
Since $\bar{\Pi}$ is a projection, directions along the fibers $\bar{\Pi}^{-1}(d)$ are also degenerate directions of $\tilde{\Omega}(s)$. It follows then, by using Cartan's formula, that the induced pre-symplectic structure $\tilde{\Omega}$ is constant along the fibers and therefore it has a well defined projection onto $\bar{\Gamma}_\text{can}$ as well. The 
 question at hand is whether the induced structure $\tilde{\Omega}(s)$ coincides with the naturally defined $\boldsymbol{\omega}(s)$ coming from the covariant phase space formalism. If that happens, namely if,
\begin{equation}
\tilde{\Omega}(s)=\boldsymbol{\omega}(s) \label{LWequivalence}
\end{equation}
we shall say that there is a precise sense of equivalence between the two formalisms. In that case, let us name both pre-symplectic structures $\bar\Pi$-equivalent (See Fig. \ref{f2}). 

We have made explicit the sense in which the two pre-symplectic structures are equivalent, namely if Eq.(\ref{LWequivalence}) is satisfied. At this stage, we should point out that in the pioneering paper by Lee and Wald,
the many-to-one mapping $\bar{\Pi}_{\text{LW}}:\Gc\to\cGc$ is a priori different from the canonical projection $\bar{\Pi}$ arising from the  3+1 decomposition, so the fiber bundle structure  induced on $\Gc$ by these mappings may in general be different: In \cite{Wald}, the degenerate directions of $\boldsymbol{\omega}$ in the whole configuration space $\mathcal{F}$ of the covariant theory are shown to be integrable and hence to define a foliation of $\mathcal{F}$.
The complete canonical phase space $\Gamma_\text{can}$ is  identified with the set of equivalence classes defined by these submanifolds of degenerate directions of $\boldsymbol{\omega}$, 
with $\Pi_{\text{LW}}:\mathcal{F}\to\Gamma_\text{can}$, the projection to the equivalence classes.
The mapping $\bar{\Pi}_{\text{LW}}:\Gc\to\cGc$ is the restriction of this projection to the space of solutions to the equations of motion.
 
By construction, the projection of $\boldsymbol{\omega}$ to the  phase space $\Gamma_\text{can}$ gives rise to a well  defined symplectic structure $\Omega_\text{LW}$. This symplectic structure and its pullback $\bar{\Omega}_\text{LW}$ to the constraint surface $\cGc$, will hence, by construction, satisfy the analogous of Eq.(\ref{LWequivalence}).
While the authors verify the equivalence of $\bar{\Omega}_\text{LW}$ with the pre-symplectic structure $\bar{\Omega}$ arising from the 3+1 decomposition for specific examples (which implies also correctly identifying their phase space of equivalence classes with initial data), the whole analysis does not consider regions with boundaries and a general discussion of equivalence is lacking.
Our Eq.(19) is asking for such a comparison, even when written on the point $s$ on the fibre. One could project down the equation to the base point in $\cGc$, and re-estate the equivalence there. Both possibilities are mathematically equivalent.

 In the recent paper [2], the authors make a strong claim of equality between symplectic structures, for generic theories and with the inclusion of boundaries. It is not clear whether their results apply to gauge theories. In what follows we shall consider several examples, and will find that even when in several cases we do have $\bar\Pi$-equivalence, we find one example for which Eq.(\ref{LWequivalence}) is violated. This counterexample is enough to invalidate $\bar\Pi$-equivalence (or correspondingly the equivalence of $\bar{\Omega}_\text{LW}$ with $\bar{\Omega}$),  as a generic property for gauge theories.
Let us now consider several concrete examples.


\section{Maxwell theory}
\label{sec:2.5}

Our first example is the Maxwell theory defined on $(\mathcal{M},  g_{ab})$, a spacetime  region $\mathcal{M}$ with boundary, $\partial\mathcal{M}=\Sigma_1\cup\Sigma_2\cup\mathcal{B}$, where $\Sigma_{1,2}$ are two (arbitrary) Cauchy surfaces and $\mathcal{B}$ is a time-like hypersurface.

We start from a covariant action, that takes the form (up to the multiplicative factor) 
\be
S_{\mathrm{M}}=\int_{\mathcal{M}} \mathbf{F}\wedge \star \mathbf{F}  
 \, ,\label{Action M}
\ee
where ${\mathbf{F}}=\md {\mathbf{A}}$ is the field strength two-form, the curvature of the $U(1)$ connection one-form ${\mathbf{A}}$ 
and  $\star \mathbf{F}$ is the Hodge dual in four dimensions of $\mathbf{F}$. Let us now see how the two Hamiltonian descriptions arise, recalling some of the results of \cite{Corichi1998}, and taking into account contributions from the boundary. 


\subsection{Covariant approach}

The variation of (\ref{Action M}) takes the form
\be
\delta S_{\mathrm{M}} 
= 2\int_{\mathcal{M}}\md{\star\mathbf{F}}\wedge\delta\mathbf{A}
 + 2 \int_{\mathcal{\partial M}}
\star \mathbf{F}\wedge\delta\mathbf{A}\, ,
\ee
resulting in the vacuum Maxwell equations in the bulk
\be
\md{\mathbf{F}}=0\ \ \text{and}\ \ \md{\star\mathbf{F}}=0\, ,\label{EOM_M}
\ee
and the boundary condition
\be
\int_{\mathcal{\partial M}}
\star \mathbf{F}\wedge\delta\mathbf{A}=0\, ,\label{BC_M}
\ee
that needs to be satisfied to have a consistent action principle.
The covariant phase space $(\Gamma_{\mathrm{CovM}},{\pmb{\omega}}_{\mathrm{M}})$ is the space of solutions of (\ref{EOM_M}), that satisfy the boundary conditions (\ref{BC_M}). We will asume that $\int_{\mathcal{B}}
\star \mathbf{F}\wedge\delta\mathbf{A}=0$, so that the  pre-symplectic structure is of the form
\be
\pmb{\omega}_{\mathrm{M}}(\delta_1,\delta_2)=4\int_\Sigma \delta_{[1}{\star\mathbf{F}}\wedge \delta_{2]}\mathbf{A}.\label{SSMCov}
\ee
It is easy to show that $\delta \mathbf{A}=\md\alpha$ are degenerate directions of ${\pmb{\omega}}_{\mathrm{M}}$.  

As shown in \cite{Corichi1998}, ${\pmb{\omega}}_{\mathrm{M}}$ can be rewritten in terms of variables defined on $\Sigma$ as,
\be
\pmb{\omega}_{\mathrm{M}}(\delta_1,\delta_2)=4\int_\Sigma \delta_{[1}\ast{(n\cdot \mathbf{F})}\wedge\delta_{2]}\mathbf{A}\, ,
\label{CovMax}
\ee
where $\ast$ denotes a three-dimensional Hodge star, and $n^a$ is unit normal to $\Sigma$. 
The normal component of $ \mathbf{F}$ is proportional to the electric field $E$ on $\Sigma$. Let us now see what is the structure of the canonical formalism.


\subsection{Canonical approach}

The first step in the Hamiltonian analysis is to make a $3+1-$decomposition of the action $S_{\mathrm{M}}$. We shall consider that ${\mathcal{M}}=\mathcal{I} \times \Sigma$, with $\mathcal{I}$ a closed interval. $\Sigma$ is a three dimensional manifold with boundary $\partial\Sigma$ that has the topology of a two sphere $S^2$.  
We introduce an everywhere timelike vector field $t^{a}$ and a ``time'' function $t$ such that hypersurfaces 
$t = {\mr const.}$ are diffeomorphic to $\Sigma$ and  $t^{a} \nabla_{a} t = 1$. 
 The kinematical symplectic structure $\Omega_{\mr{M}}$ is of the form
\be
\Omega_{\mr{M}}(\delta_1,\delta_2)= 2 \int_{\Sigma} \delta_{[1}{\mathbf{P}}_\varphi \wedge \delta_{2]}\varphi  +
\delta_{[1}\mathbf{P}\wedge\delta_{2]}\mathbf{A}\, ,
\ee
where $\varphi =t\cdot\mathbf{A}$. 
The theory has two first class constraints:
\ba
{\mathbf{C}}_\varphi &:=& {\mathbf{P}}_{\varphi}\approx 0\, ,\\
C &:=& \ast{\md {\mathbf{P}}} \approx 0\, ,
\ea
where the Gauss constraint $C$ is obtained from the consistency condition of the primary contraint ${\mathbf{C}}_\varphi$. There are no further constraints.

In this case the symplectic structure  $\Omega_{\mr{M}}$ does not have any boundary contribution, so that the variation of the allowed phase space functionals cannot have any boundary terms, that impose some boundary conditions. 
The Hamiltonian $H_{\mathrm{M}}$ and the corresponding boundary conditions needed for its differentiability are given in \cite{Corichi1998}.

In order to compare the covariant and the canonical symplectic structure, we can rewrite $\Omega_{\mr{M}}$, using the definition of the canonical momenta.
where the canonical momenta are given by
\ba
{\mathbf{P}}_{\varphi} &=&0\, ,\\
{\mathbf{P}} &=& 
2 \ast{(n\cdot \mathbf{F})}\, .
\ea
As a result, we obtain
\be
\tilde{\Omega}_{\mr{M}}(\delta_1,\delta_2)= 4 \int_{\Sigma} 
\delta_{[1}\ast{(n\cdot \mathbf{F})}\wedge\delta_{2]}\mathbf{A}
\, ,
\ee
which is $\bar\Pi$-equivalent to the covariant symplectic structure ${\pmb{\omega}}_{\text{M}}(\delta_1,\delta_2)$ (\ref{CovMax}). 

Let us now see how one can understand equivalence at the level of reduced phase spaces \cite{Corichi1998}. In the covariant approach, the reduced phase space $\hat{\Gamma}_{\text{cov}}$ can be characterized by the equivalence class of
4D connections $[\mathbf{A}]$, or by the curvature tensor $\mathbf{F}$ satisfying both of Maxwell's equations (\ref{EOM_M}). Equivalently, given a hypersurface $\Sigma$, $\mathbf{F}$ can be decomposed into a pair $(B,E)$, namely the electric and magnetic field, such that $E$ satisfies the initial data constraints.
In turn, this characterization is equivalent to the canonical one, where the reduced phase space $\hat{\Gamma}_{\text{can}}$ is given by pairs $([\mathbf{A}],E)$ where the equivalence class in now of spatial $U(1)$ connections, and $E$ satisfies Gauss' law.


\section{Pontryagin theory}
\label{sec:3}

As another example of a gauge theory defined in a region with boundary, let us consider a topological theory, defined in the bulk, namely 
the $U(1)$ Pontryagin theory on the 4D spacetime region ${\mathcal{M}}$, with boundary.
This section has two parts. In the first one, we perform the covariant Hamiltonian analysis of the theory and in the second one, we recall some results of the corresponding canonical analysis \cite{CV-P}.  

The Pontryagin action for the Abelian theory is
\begin{equation}\label{Pontryagin}
S_{\mr{P}} =  \theta \int_{\mathcal{M}} {\mathbf{F}}\wedge {\mathbf{F}}
\, ,
\end{equation}
where $\theta$ is an arbitrary real parameter.


\subsection{Covariant approach}

The variation of (\ref{Pontryagin}) only has a boundary contribution, since the bulk equation of motion, $\md\mathbf{F}=0$, is trivially satisfied, 
\be
\delta S_{\mr{P}}= 2 \theta  \int_{\partial\mathcal{M}}  \mathbf{F}\wedge\delta {\mathbf{A}}\, ,\label{VarPont1}
\ee
where
$\partial{\mathcal{M}} =\Sigma_1\cup \Sigma_2\cup {\mathcal{B}}$, with ${\mathcal{B}}=I\times\partial\Sigma$. This term should vanish in order to have well defined variational principle, even in this case when there are no equations of motion in the bulk. Since $\delta{\mathbf{A}}=0$ on $\Sigma_1$ and $\Sigma_2$, we see that the condition
\be
 \int_{{\mathcal{B}} } {\mathbf{F}}
 \wedge\delta {\mathbf{A}}=0\, ,\label{BoundCond1}
\ee
defines boundary conditions of the theory. One possibility is that
\be
\mathbf{F}=0\, , \ \ \text{on}\ \ {\mathcal{B}},\label{FZero}
\ee
but (\ref{BoundCond1}) can be fulfilled even without (\ref{FZero}), as is the case for perfect conductor boundary conditions \cite{CV-M+P}. 
The covariant phase space $(\Gamma_{\text{Pcov}} , \pmb{\omega}_{\mr{P}})$ is then defined as a space of {\it all} connections such that (\ref{BoundCond1}) holds.  

The corresponding symplectic structure ${\pmb{\omega}}_{\mr{P}}$ must vanish identically, since (\ref{Pontryagin}) is a topological term. Indeed,  (\ref{VarPont1}) can be rewritten as
\be
\ed S_{\mr{P}} (\delta):=\delta S_{\mr{P}}=  \theta  \int_{\partial\mathcal{M}} \delta ( \mathbf{F}\wedge {\mathbf{A}})\, .\label{VarPont2}
\ee
As a consequence, as we have seen in Sec.\ref{sec:2} the symplectic current $J(\delta_1,\delta_2)$ vanishes,
as well as the symplectic structure on the covariant phase space of the theory 
\be
{\pmb{\omega}}_{\mr{P}}(\delta_1,\delta_2)=0\, .
\ee
This implies that every direction $\delta\mathbf{A}\in T\Gamma_{\text{Pcov}}$ is a degenerate one. The reduced phase space $(\hat{\Gamma}_{\text{Pcov}} , \hat{\pmb{\omega}}_{\mr{P}})$, is then trivial, in the sense that it does not have local degrees of freedom. There can only be global boundary degrees of freedom, when the boundary has non trivial cohomology.
 

\subsection{Canonical approach}

The first step in the Hamiltonian analysis is to make a $3+1-$decomposition of the action. As before, we introduce an everywhere timelike vector field $t^{a}$ and a ``time'' function $t$ such that $t^{a} \nabla_{a} t = 1$. 
Then, the canonical form of $S_{\mr{P}}$ is
\be\label{PontryaginCan}
S_{\text{Pcan}}=2 \theta \int_I \md t \int_\Sigma (t\cdot \mathbf{F})\wedge \mathbf{F}
=2 \theta \int_I \md t \int_\Sigma  [\mathcal{L}_t\mathbf{A}-\md\varphi]\wedge \mathbf{F} \, .
\ee
From (\ref{PontryaginCan}) we can read off the form of the kinematical symplectic structure 
\be
\Omega_{\mr{P}}(\delta_1,\delta_2)= 2 \int_\Sigma \delta_{[1}{\mathbf{\Pi}}_{\text{P}\varphi} \wedge \delta_{2]}\varphi  +
\delta_{[1}{\mathbf{\Pi}}_{\text{P}}\wedge\delta_{2]}\mathbf{A}\, ,
\ee
and the corresponding proposal for the canonical Hamiltonian, 
$H_{\mr{P}} =  \int_\Sigma \md\varphi\wedge{\mathbf{\Pi}}_{\text{P}}$.
There are 
four primary constraints
\ba
{\mathbf{C}}_{\text{P}\varphi} &:=& {\mathbf{\Pi}}_{\text{P}\varphi}\approx 0\, ,\\
{\mathbf{C}}_{\text{P}} &:=& {\mathbf{\Pi}}_{\text{P}}- 2\theta\, \mathbf{F}\, \approx 0\, .
\ea
Following Dirac's algorithm we define the total Hamiltonian as
$
H_{\mr{PT}} =  \int_\Sigma (\md\varphi\wedge{\mathbf{\Pi}}_{\text{P}}
+ \mathbf{u}\wedge {\mathbf{C}}_{\text{P}} + \mathbf{v}\wedge {\mathbf{C}}_{\text{P}\varphi})    
$,
where $\mathbf{u}$ and $\mathbf{v}$ are the corresponding smearing one-form and function. 
The total Hamiltonian defines the evolution via its corresponding Hamiltonian vector fields (HVF) $X_H$, given by 
\be
\ed H_{\mr{PT}}=\Omega_{\mr{P}}(\cdot ,X_H)\, .
\ee
Since ${\Omega}_{\mr{P}}$ only has a bulk contribution, it follows that $\ed H_{\mr{PT}}$ cannot have any boundary terms, leading to two conditions
$
\int_{\partial\Sigma}{\mathbf{\Pi}}_{\text{P}}\wedge \delta\varphi =0
$
and
$
\int_{\partial\Sigma}\mathbf{u}\wedge \mathbf{\delta A}=0
$.

It is known that all of primary constraints are first class and that there are no new secondary class constraints. If we denote the HVF corresponding to the four FC contraints $({\mathbf{C}}_{\text{P}\varphi} , {\mathbf{C}}_{\text{P}a} )$ as $Z_i$ \footnote{In order to have well defined HVF $Z_i$ we need to impose certain restrictions on the smearing functions on $\partial\Sigma$, as shown in \cite{CV-P}.}, then the pullback of $\Omega_{\mr P}$ to the constraint surface, $\bar{\Gamma}_{\text{Pcan}}$, 
is degenerate, 
$
{\bar{\Omega }}_{\mr P}(Y,Z_i)=0
$,
for every $Y\in T{\bar\Gamma}_{\text{Pcan}}$. In particular
\be
{\bar{\Omega}}_{\mr P}(Z_i,Z_j)=0\, .
\ee
Since HVF $Z_i$ span the four-dimensional $T{\bar\Gamma}_{\text{Pcan}}$,  the  corresponding pullback of the symplectic structure ${\bar{\Omega }}_{\mr P}$ is trivial
\be
{\bar{\Omega }}_{\mr P} (X,Y)=0\, ,
\ee
for every $X,Y \in T{\bar\Gamma}_{\text{Pcan}}$. 

It is at this level where we can ask for the equivalence between covariant and canonical pre-symplectic structures. The first observation is that the $\bar\Pi$-map does not exist. 
To begin with, the space  $\Gamma_{\text{Pcov}}$ of solutions is the full space of field configurations $\cal{F}$. If we induce initial data
from any of such configurations on a hypersurface $\Sigma_0$, this data will {\it{not}} in general satisfy the four first class constraints of the canonical theory. Thus, if the $\bar\Pi$-map does not exist, the two descriptions can not be possibly $\bar\Pi$-equivalent. They are equivalent, however, at the level of reduced phase spaces. We have seen that the covariant pre-symplectic structure is trivial, so every direction is a gauge direction. In the canonical case, the  induced structure ${\bar\Omega}_{\mr{MP}}$ is also trivial in
$\bar{\Gamma}_{\text{Pcan}}$, so all tangent directions to it are also gauge directions. The final description is that in the canonical theory, the reduced phase space 
$\hat{\Gamma}_{\text{Pcan}}$ has no local degrees of freedom either.



\section{Chern-Simons theory on the boundary}
\label{sec:4}

It is well known that  the Pontryagin action, being the integral of a total derivative, can be rewritten as an integral of the Chern-Simons action over the boundary $\partial{\mathcal{M}}$ of the spacetime region under consideration. In \cite{CV-P} the Pontryagin theory on  ${\mathcal{M}}$ and the Chern-Simons term on $\mathcal{B}=I\times\partial\Sigma$ were compared within the canonical Hamiltonian description and shown to be equivalent. In this section we shall perform the  covariant analysis in the first part. In the second one, we recall the results of the canonical theory, and compare them.

\subsection{Covariant approach}

The action given by the Pontryagin term on $\mathcal{M}$ is equivalent to Chern-Simons action on  $\partial{\mathcal{M}}$, We shall consider the Chern-Simons theory on $\mathcal{B}=\mathcal{I}\times\partial\Sigma$ instead, since we want to compare the covariant Hamiltonian theory with the canonical analysis of \cite{CV-P}. This choice might
have some consequences as we shall discuss below. The starting action has the form,
\be
S_{\mr{CS}}=\theta\, \int_{\mathcal{B}} \mathbf{A}\wedge\mathbf{F}\, .
\ee
Then,
\be
\ed S_{\mr{CS}} (\delta):=\delta S_{\mr{CS}}=2\theta\, \int_{\mathcal{B}}\mathbf{F}\wedge\delta \mathbf{A} - \theta\, \int_{\partial\mathcal{B}} \mathbf{A}\wedge\delta\mathbf{A}\, ,
\ee
where $\partial\mathcal{B}=\partial\Sigma_1\cup\partial\Sigma_2$. The corresponding equations of motion are given by $\mathbf{F}=0$, on ${\mathcal{B}}$. Note that the last integral vanishes, since $\delta\mathbf{A}=0$ on $\Sigma_1$ and $\Sigma_2$.   
The symplectic current $J_{\mr{CS}}(\delta_1,\delta_2)$ is obtained from
\be
\ed^2 S_{\mr{CS}}(\delta_1,\delta_2) := \int_{\partial\mathcal{B}}J_{\mr{CS}}(\delta_1,\delta_2)= -2\theta\,\int_{\partial\mathcal{B}} \delta_1\mathbf{A}\wedge\delta_2\mathbf{A} =   0\, ,
\ee
resulting in the conserved pre-symplectic structure
\be
{\pmb{\omega}}_{\mr{CS}}(\delta_1,\delta_2)= -2\theta\, \int_{\partial\Sigma}\delta_1\mathbf{A}\wedge\delta_2\mathbf{A}\, .
\ee
Note that it does not identically vanish, as in the case of the Pontryagin term. Nevertheless, since $\mathbf{F}=0$ on $\mathcal{B}$ it implies that, locally, the allowed variations are of the form $\delta\mathbf{A}=\md\alpha$ on $\partial\Sigma$.\footnote{Recall that we are assuming $\partial{\Sigma}$ having a vanishing first cohomology group.} Then, for such generic tangent vectors to $\Gc$ we have
\be
{\pmb{\omega}}_{\mr{CS}}(\delta_1,\delta_2)= -2\theta\, \int_{\partial\Sigma}
\md\alpha_1\wedge\md\alpha_2=0\, .
\ee
Thus, every tangent vector to $\Gc$ is a degenerate direction of ${\pmb{\omega}}_{\mr{CS}}$, which means that the space of orbits defined by the gauge directions is trivial
leading, as expected, to the same result as in the covariant Pontryagin theory on $\mathcal{M}$. Let us now recall the canonical description of the CS-theory \cite{CV-P}.


\subsection{Canonical approach}
The canonical action is
\be
S_{\mr{CScan}}=-\theta \int_I \md t \int_{\partial\Sigma}  [\mathcal{L}_t\mathbf{A}\wedge\mathbf{A} + \varphi\wedge \mathbf{F} ]\, ,
\ee
from which we can read off the form of the kinematical symplectic structure 
\be
\Omega_{\mr{CS}}(\delta_1,\delta_2)= 2 \int_{\partial\Sigma} \delta_{[1}{\mathbf{\Pi}}_{\mr{CS}\varphi} \wedge \delta_{2]}\varphi  +
\delta_{[1}{\mathbf{\Pi}}_{\mr{CS}}\wedge\delta_{2]}\mathbf{A}\, ,
\ee
and the corresponding proposal for the canonical Hamiltonian, 
$
H_{\mr{CS}} = \theta \int_{\partial\Sigma} \varphi\wedge\mathbf{F}\, .
$
There are 
three primary constraints
\ba
{\mathbf{C}}_{\mr{CS}\varphi} &:=& {\mathbf{\Pi}}_{\mr{CS}\varphi}\approx 0\, ,\\
{\mathbf{C}}_{\mr{CS}} &:=& {\mathbf{\Pi}}_{\mr{CS}}  +\theta\,  \mathbf{A}\, \approx 0\, .
\ea
The HVF $\delta_C$ corresponding to the smeared constraint
$
{\mathbf{C}}_{\mr{CS}}[\mathbf{u}]=\int_{\partial\Sigma} \mathbf{u}\wedge{\mathbf{C}}_{\mr{CS}}\, ,
$
is obtained from
\be
\ed {\mathbf{C}}_{\mr{CS}}[\mathbf{u}] (\delta) = \Omega_{\mathrm{CS}} (\delta,\delta_C)\, .
\ee
It turns out that
$
\Omega_{\mathrm{CS}} (\delta_{C_1},\delta_{C_2})=2\theta\, \int_{\partial\Sigma} \mathbf{u_1}\wedge\mathbf{u_2}\ne 0\, ,
$
so that there are two primary second class constraints ${\mathbf{C}}_{\mr{CS}a}\approx 0$. 
The total Hamiltonian is
\be
H_{\mr{TCS}} = \int_{\partial\Sigma} [\theta\, \varphi\wedge\mathbf{F} + \mathbf{v}\wedge {\mathbf{C}}_{\mr{CS}} + w\wedge {\mathbf{C}}_{\mr{CS}\varphi}] \, ,
\ee
and its corresponding HVF $\delta_H$ is defined as
\be
\ed H_{\mr{TCS}} (\delta) = \Omega_{\mathrm{CS}} (\delta,\delta_H)\, .
\ee
The consistency condition for the primary constraint ${\mathbf{C}}_{\mr{CS}\varphi} [w_1]$ leads to a secondary constraint
\be
\mathcal{L}_t {\mathbf{C}}_{\mr{CS}\varphi} [w_1] :=
\Omega_{\mathrm{CS}}  (\delta_{{\mathbf{C}}_\varphi}, \delta_H)=
\int_{\partial\Sigma} w_1\wedge \mathbf{F}\, \ \ \Rightarrow\ \ \mathbf{F}=0\ \
\mathrm{on}\ \ \partial\Sigma\, ,
\ee
while from the consistency condition for ${\mathbf{C}}_{\mathrm{CS}}[\mathbf{u}]$ we obtain the multiplier $\mathbf{v}$,
\be
\mathcal{L}_t {\mathbf{C}}_{\mathrm{CS}} [\mathbf{u}] :=
\Omega_{\mathrm{CS}}  (\delta_{{\mathbf{C}}}, \delta_H) =\theta\, \int_{\partial\Sigma} (\md\varphi - 2\mathbf{v})\wedge\mathbf{u}=0
\, \ \ \Rightarrow\ \ \mathbf{v}=\frac{1}{2}\md\varphi \ \
\mathrm{on}\ \ \partial\Sigma\, .
\ee
There are no tertiary constraints. 
Let us define $\Gamma_{\text{CSD}}\subset{\Gamma}_{\text{CScan}}$ as a hypersurface where ${\mathbf{C }}_{\mr{CS}}\approx 0$. Then, the pullback of the symplectic structure to $\Gamma_{\mr{CSD}}$ is non degenerate,
\be
\Omega_{\mathrm{CSD}} (\delta_1,\delta_2)=2\int_{\partial\Sigma}  -\theta\, \delta_1\mathbf{A}\wedge \delta_2\mathbf{A} +  \delta_{[1}{\mathbf{\Pi}}_{\mr{CS}\varphi}\wedge\delta_{2]}\varphi \, .
\ee
 $\Gamma_{\text{CSD}}$ is four dimensional (per point), and there are still two first class constraints
\be
\mathbf{\Pi}_{\mr{CS}\varphi}\approx 0\, , \ \ \mathbf{F}\approx 0\, .
\ee
These constraints define  ${\bar{\Gamma}}_{\text{CScan}}\subset\Gamma_{\text{CSD}}$, whose tangent vectors are such that $\delta {\mathbf{\Pi}}_{\mr{CS}\varphi}\approx 0$ and locally $\delta\mathbf{A}\approx\md\mathbf{\alpha}$. The resulting degenerate pre-symplectic structure is trivial
\be
\bar{\Omega}_{\mathrm{CSD}}(X,Y)=0\, ,
\ee
for every $X,Y\in T{\bar{\Gamma}}_{\text{CScan}}$.  Note that in this case the $\bar\Pi$-map is well defined, since we do have covariant equations of motion. Furthermore, just as in the covariant description, at the pre-symplectic level both  ${\pmb{\omega}}_{\mr{CS}}$
and  $\bar{\Omega}_{\mathrm{CSD}}$ act trivially on tangent vectors to the corresponding spaces. 
 Thus, we can conclude that $\bar\Pi$-equivalence is satisfied in a trivial way.


\section{Maxwell-Pontryagin theory}
\label{sec:5}

In this section we consider the Maxwell-Pontryagin theory in 4D. We start from the covariant action and develop the two Hamiltonian descriptions. In the first part we consider the covariant theory, and in the second we recall the canonical analysis of \cite{CV-M+P}.

The starting point is the covariant action, that is a sum of the action for the Maxwell theory and the Pontryagin term, and takes the form (up to the multiplicative factor) 
\be
S_{\mathrm{MP}}=\int_{\mathcal{M}} \mathbf{F}\wedge \star \mathbf{F} + \theta\,
\mathbf{F}\wedge\mathbf{F}\, ,\label{Action M+P}
\ee
where $\star \mathbf{F}$ is the Hodge dual in four dimensions of $\mathbf{F}$.


\subsection{Covariant approach}

The variation of (\ref{Action M+P}) takes the form
\be
\delta S_{\mathrm{MP}}=2\int_{\mathcal{M}}
 \md{\star\mathbf{F}}\wedge \delta\mathbf{A}+ 2 \int_{\mathcal{\partial M}}
(\star \mathbf{F}+\theta\,\mathbf{F})\wedge\delta\mathbf{A}\, ,
\ee
resulting in the vacuum Maxwell equations in the bulk
\be
\md{\mathbf{F}}=0\ \ \text{and}\ \ \md{\star\mathbf{F}}=0\, ,\label{EOM_MP}
\ee
and the boundary condition
\be
\int_{\mathcal{\partial M}}
(\star \mathbf{F}+\theta\,\mathbf{F})\wedge\delta\mathbf{A}=0\, .\label{BC_MP}
\ee
The covariant phase space $(\Gamma_{\mathrm{MPcov}},{\pmb{\omega}}_{\mathrm{MP}})$ is the space of the solutions of (\ref{EOM_MP}), that satisfy the $\theta$-dependant boundary conditions (\ref{BC_MP}) (which makes it different to the pure Maxwell theory). Furthermore, we shall impose that there is no
symplectic leakage across the boundary, namely $\int_{\mathcal{B}}\delta_{[1}{\star\mathbf{F}}\wedge \delta_{2]}\mathbf{A}=0$.\footnote{Note that, as pointed out in the preliminaries, there could be boundary conditions that render the symplectic current exact on $\mathcal{B}$, in which case we would have a contribution to the pre-symplectic structure from the boundary of $\Sigma$.} This is a necessary condition for the existence of the conserved pre-symplectic structure of the form
\be
{\pmb\omega}_{\mathrm{MP}}(\delta_1,\delta_2)=4\int_\Sigma \delta_{[1}{\star\mathbf{F}}\wedge \delta_{2]}\mathbf{A}={\pmb\omega}_{\mathrm{M}}(\delta_1,\delta_2)\, .\label{SSMPCov}
\ee
As we have shown in Sec.\ref{sec:2} the Pontryagin term does not contribute to ${\pmb\omega}_{\mathrm{MP}}$. It is easy to show that $\delta \mathbf{A}=\md\alpha$ are degenerate directions of ${\pmb\omega}_{\mathrm{MP}}$, which are precisely the gauge directions of the Maxwell theory. 

As shown in \cite{Corichi1998}, ${\pmb\omega}_{\mathrm{MP}}$ can be rewritten in terms of variables defined on $\Sigma$
\be
{\pmb\omega}_{\mathrm{MP}}(\delta_1,\delta_2)=4\int_\Sigma \delta_{[1}\ast{(n\cdot \mathbf{F})}\wedge\delta_{2]}\mathbf{A}\, ,
\ee
where $\ast$ denotes a three-dimensional Hodge star, and $n^a$ is unit normal to $\Sigma$.

Let us now consider the canonical description, starting from the same action.


\subsection{Canonical approach}

We start with the $3+1$ decomposition  of the covariant action $S_{\mathrm{MP}}$ from which we arrive to the corresponding canonical action \cite{CV-M+P}. The kinematical symplectic structure is of the form
\be
\Omega_{\mr{MP}}(\delta_1,\delta_2)= 2 \int_{\Sigma} \delta_{[1}{\mathbf{\Pi}}_\varphi \wedge \delta_{2]}\varphi  +
\delta_{[1}\mathbf{\Pi}\wedge\delta_{2]}\mathbf{A}\, .
\ee
The symplectic structure  does not have any boundary contribution, so the variation of the allowed phase space functionals cannot have any boundary terms. This condition  imposes some boundary conditions for the fields. 
The Hamiltonian $H_{\mathrm{MP}}$ and the corresponding ($\theta$-dependent) boundary conditions needed for its differentiability are given in \cite{CV-M+P}. 


The theory has two first class constraints:
\ba
{\mathbf{C}}_\varphi &:=& {\mathbf{\Pi}}_{\varphi}\approx 0\, ,\\
C &:=& \ast{\md {\mathbf{\Pi}}} \approx 0\, ,\label{ConstraintsMP}
\ea
where the Gauss constraint $C$ is obtained from the consistency condition of the primary contraint ${\mathbf{C}}_\varphi$. There are no further constraints.

In order to compare the covariant and the canonical symplectic structure, we can rewrite $\Omega_{\mr{MP}}$, using the definition of the canonical momenta,
\ba
{\mathbf{\Pi}}_{\varphi} &=&0\, ,\\
{\mathbf{\Pi}} &=& 
2\ast{(n\cdot \mathbf{F})}- \theta\,\mathbf{F}\, .
\ea
As a result, we obtain
\be
{\tilde\Omega}_{\mr{MP}}(\delta_1,\delta_2)= 4 \int_{\Sigma} 
\delta_{[1}\ast{(n\cdot \mathbf{F})}\wedge\delta_{2]}\mathbf{A}
-2\theta\int_{\partial{\Sigma}}\delta_{1}\mathbf{A}\wedge\delta_{2}\mathbf{A}
\, ,
\ee
that, apart from the covariant pre-symplectic structure of the Maxwell theory, has an additional Chern-Simons boundary contribution. Since $\mathbf{F}\ne 0$ on $\partial\Sigma$, the boundary term in ${\tilde\Omega}_{\mr{MP}}$ does not vanish. 

Thus, we see that in this case there is no $\bar\Pi$-equivalence between the pre-symplectic structures. This is precisely the counterexample that shows that 
$\bar\Pi$-equivalence is not a generic feature of gauge theories.

Let us now consider the final model of this manuscript, namely Maxwell+Chern-Simons, that is known to be equivalent to the Maxwell-Pontryagin theory of this section. 


\section{Maxwell + Chern-Simons}
\label{sec:6}

Let us now analyze the case where the Pontryagin term on $\mathcal{M}$ is replaced by  Chern-Simons term on the boundary $\mathcal{B}$, 
\be
S_{\mathrm{MCS}}=\int_{\mathcal{M}} \mathbf{F}\wedge \star \mathbf{F} +\theta\, \int_{\mathcal{B}} \mathbf{A}\wedge\mathbf{F}\, .\label{Action M+CS}
\ee


\subsection{Covariant Approach}

Now, the variation of the action is of the form
\be
\delta S_{\mathrm{MCS}} = 2\int_{\mathcal{M}}
 \md{\star\mathbf{F}}\wedge \delta\mathbf{A} + 2 \int_{\partial\mathcal{M}}
\star \mathbf{F}\wedge\delta\mathbf{A}+2\theta\,\int_{\mathcal{B}}\mathbf{F}\wedge\delta\mathbf{A}
 - \theta\int_{\partial\mathcal{B}} \mathbf{A}\wedge\delta\mathbf{A}\,  ,
\ee
again leading to the Maxwell equations in the bulk. 
In the third integral we can replace $\mathcal{B}$ by $\partial\mathcal{M}=\mathcal{B}\cup\Sigma_1\cup\Sigma_2$, since $\delta\mathbf{A}=0$ on $\Sigma_{1,2}$, being also the reason for the vanishing of the last integral. 
In this way we obtained the same condition as in the Maxwell-Pontryagin theory, that now could be interpreted in two ways: as a boundary condition on $\partial\mathcal{M}$ or as the equation of motion on $\mathcal{B}$
\[
\star \mathbf{F}+\theta\,\mathbf{F}=0\, , \ \ {\text{on}}\ \  \mathcal{B}\, .
\]
The corresponding symplectic structure now has a boundary term
\be
{\pmb{\omega}}_{\mathrm{MCS}}(\delta_1,\delta_2)=2\int_\Sigma \delta_{[1}{\star\mathbf{F}}\wedge \delta_{2]}\mathbf{A} - 2\,\theta \int_{\partial\Sigma}\delta_1\mathbf{A}\wedge\delta_2\mathbf{A} \, .\label{SSMCSCov}
\ee
Again, the boundary term in the symplectic structure does not vanish, as it was the case in Chern-Simons theory.



\subsection{Canonical Approach}

Here we shall recall the main results from \cite{CV-M+P}. After performing the 3+1 decomposition of the action (\ref{Action M+CS}), we can arrive 
at the kinematical symplectic structure, that now has a boundary term and is of the form
\be
\Omega_{\mr{MCS}}(\delta_1,\delta_2)= 2 \int_{\Sigma} \delta_{[1}{\mathbf{P}}_\varphi \wedge \delta_{2]}\varphi  +
\delta_{[1}\mathbf{P}\wedge\delta_{2]}\mathbf{A} +
2 \int_{\partial\Sigma} \delta_{[1}{\mathbf{P}}_{\varphi^\partial} \wedge \delta_{2]}\varphi^{\partial}  +
\delta_{[1}{\mathbf{P}}_{\partial}\wedge\delta_{2]}{\mathbf{A}}^{\partial}
\, ,
\ee
where ${\mathbf{A}}^{\partial}$ is the pullback of ${\mathbf{A}}$ to ${\partial\Sigma}$, ${\varphi^\partial}$ is $\varphi$ evaluated on ${\partial\Sigma}$. Furthermore, ${\mathbf{P}}_{\partial}$ and ${\mathbf{P}}_{\varphi^\partial}$ are their corresponding canonical momenta that are not related to the canonical momenta in the bulk.
The bulk part of the symplectic structure corresponds to the Maxwell theory, while the boundary contribution comes from the Chern-Simons term on the boundary. 

As in the case of Maxwell-Pontryagin, this theory has two first class constraints in the bulk:
\ba
{\mathbf{C}}_\varphi &:=& {\mathbf{P}}_{\varphi}\approx 0\, ,\\
C &:=& \ast{\md {\mathbf{P}}} \approx 0\, ,\label{ConstraintsMCSbulk}
\ea
where, as before, the Gauss constraint $C$ is obtained from the consistency condition of the primary contraint ${\mathbf{C}}_\varphi$. 

There are also constraints on the boundary, all of them are primary, given by 
\ba
{\mathbf{C}}_{\varphi^\partial} &:=& {\mathbf{P}}_{\varphi^\partial}\approx 0\, ,\\
{\mathbf{C}}_{\partial} &:=& {\mathbf{P}}_{\partial} + \theta {\mathbf{A}}^{\partial}\approx 0\, .\label{ConstraintsMCSboundary}
\ea
The consistency condition for ${\mathbf{C}}_{\varphi^\partial}$ leads to the boundary condition for the bulk variables, 
$
\ast{(r\cdot \mathbf{P})}-\theta \mathbf{F}=0\,\vert_{\partial\Sigma}\, ,
$
where $r^a$ is the exterior unit normal to $\partial\Sigma$, such that $r^a n_a=0$. 
These are the same boundary conditions that were obtained in the Maxwell-Pontryagin theory, as one of the necessary conditions for the differentiability of the Hamiltonian \cite{CV-M+P}. On the other hand, the consistency conditions for ${\mathbf{C}}_{\partial}$ lead to the determination of the corresponding multipliers in the boundary term of the Hamiltonian of the theory, due to the second class nature of these constraints. 


As in the previous case, we can compare the covariant and the canonical pre-symplectic structures, using the definition of the canonical momenta,
given by
\ba
{\mathbf{P}}_{\varphi} &=&0\, ,\\
{\mathbf{P}} &=& 
2\ast{(n\cdot \mathbf{F})}\, ,\\
{\mathbf{P}}_{\varphi^\partial} &=& 0\, ,\\
{\mathbf{P}}_{\partial}&=& -\theta {\mathbf{A}}^{\partial}\, .
\ea
The resulting pre-symplectic structure takes the form
\be
{\tilde{\Omega}}_{\mr{MCS}}(\delta_1,\delta_2)= 4 \int_{\Sigma} 
\delta_{[1}\ast{(n\cdot \mathbf{F})}\wedge\delta_{2]}\mathbf{A} -
2 \theta \int_{\partial\Sigma}
\delta_1\mathbf{A}\wedge\delta_2{\mathbf{A}}
\, ,
\ee
where, in the second integral we have written $\mathbf{A}$, instead of ${\mathbf{A}}^{\partial}$, in order to simplify the notation. In this case we can see there is 
$\bar\Pi$-equivalence between  the pre-symplectic structures (as opposed to the Maxwell-Pontryagin case).


\section{Discussion and Conclusions}
\label{sec:7}

The main objective of this manuscript is to address the issue of equivalence between covariant and canonical Hamiltonian descriptions of gauge field theories, when boundaries are present.
Notwithstanding already available efforts to tackle this issue, we have approached the problem from a natural `canonical perspective'.  That is, we have followed standard procedures to both Hamiltonian formalisms and have compared them using a natural notion of equivalence (refered to as $\bar\Pi$-equivalence) between them. 
This notion seems not to be fully coincident with previous studies to the question at hand (see, for instance, \cite{Wald,Espanolitos}).  The main question we have tried to answer is motivated from  physical considerations: How can we be assured that the physical predictions of both approaches can be regarded as equivalent? 
At first, we formulated the problem within the available formalisms, and came forward with a proposal for equivalence. The main idea is that we should compare the two formalisms at the level of pre-symplectic structures.  That is, before taking any quotient by gauge directions.  This is the context within which we prescribed our notion of equivalence.
Next, we considered several examples of gauge theories that have already been studied in the literature, in order to revisit them from new vistas. This novel perspective has allowed us to approach  some well known systems.

We started by considering Maxwell theory, where it is easy to show that both descriptions are $\bar\Pi$-equivalent. Next, we analysed the Pontryagin term. Here, due to the topological nature of the action, we have somewhat `singular' formulations, which lie outside the realm of comparison. One can see, though, that both descriptions become equivalent at the level of (trivial) reduced phase spaces. Next, we considered the Chern-Simons theory induced at the boundary by the Pontryagin term. In this case, one can show that both formulations are equivalent. The Maxwell-Pontryagin theory was then analysed. Here we found that both descriptions are indeed {\emph {inequivalent}}; in the covariant approach there is only contribution to the pre-symplectic structure from the Maxwell term, while in the canonical analysis we have, apart from the Maxwell contribution, a $\theta$-dependent {\emph{boundary contribution}}. This results shows that $\bar\Pi$-equivalence is {\emph{not}} a generic feature of gauge field theories. Finally, we analysed Maxwell+Chern-Simons theory. In this case, there is a contribution to the pre-symplectic structure that arises from both descriptions, so they turn out to be  $\bar\Pi$-equivalent.

In this manuscript, the examples we considered can be regarded as `pure gauge theories' in the sense that the only dynamical variable is a $U(1)$ connection
$\mathbf{A}$. There are, however, other gauge theories (as defined by having FCC or degenerate directions in the pre-symplectic structure) in regions with boundaries, where both methods have been compared  (see, for instance \cite{merced-perez,barnich,review}). The most notable example is the treatment of Isolated Horizons (IH), that is, generalizations of Killing horizons to model black holes in equilibrium. These systems have been treated both in the canonical formalism \cite{Abhay1,CRV-2} and in the covariant approach \cite{afk,CRV-1}. By comparing these results we can immediately see that there is an important difference between them \cite{CRV-Next}. The most salient feature appears already within the theory in the vacuum. Here one can see that the boundary contribution to the pre-symplectic structure in the covariant theory is entirely different from the (standard) contribution to the canonical description \cite{Abhay1,CRV-2}. Even more, when coupling to the Maxwell field, there is a contribution on the IH to the pre-symplectic structure in the covariant approach \cite{afk}, while there is none in the canonical one \cite{Abhay1}.

In this contribution we have shown that a natural definition for equivalence, when comparing standard covariant and canonical approaches fails to be generically satisfied. Can we conclude, therefore, that both methods are not equivalent? Our viewpoint is that we do need a deeper understanding of the reasons for equivalence (or non-equivalence). Or, perhaps, a refined notion of equivalence, different to the one proposed here and in the literature \cite{Wald,Espanolitos}. We believe that a satisfactory answer to this issue is highly relevant, since both approaches might lead to different physical predictions for theories of interest (as they do, for example, when dealing with BH entropy). We can only hope that this contributions might  provide a motivation for further studies.


\begin{acknowledgments}
This work was in part supported by a CIC-UMSNH grant.
\end{acknowledgments}


\end{document}